\RequirePackage{fix-cm}
\documentclass{svjour3}                     
\smartqed  
\usepackage{graphicx}
\usepackage{soul}
\usepackage{soul}
\usepackage[hyphens]{url}
\usepackage[hidelinks]{hyperref}
\usepackage[utf8]{inputenc}
\usepackage[small]{caption}
\usepackage{natbib}
\usepackage{subcaption}
\usepackage{graphicx}
\usepackage{amsmath}
\usepackage{booktabs}
\usepackage{listings}
\usepackage{xcolor}
\begin{document}
\title{Bin2vec: Learning Representations of Binary Executable Programs for Security Tasks}


\author{
Shushan Arakelyan\and
Christophe Hauser\and
Erik Kline\and
Aram Galstyan}

\institute{S.Arakelyan \and S.Arasteh \and C.Hauser \and E.Kline \and A.Galstyan \at
            Information Sciences Institute, University of Southern California \\
            \email{\{shushana, arasteh\}@usc.edu, \{hauser, kline, galstyan\}@isi.edu}
}

\date{}

\maketitle

Corresponding author: Shushan Arakelyan

\begin{abstract}
Tackling binary program analysis problems has traditionally implied manually defining rules and heuristics, a tedious and time consuming task for human analysts. In order to improve automation and scalability, we propose an alternative direction based on distributed representations of binary programs with applicability to a number of downstream tasks.
We introduce Bin2vec, a new approach leveraging Graph Convolutional Networks (GCN) along with computational program graphs in order to learn a high dimensional representation of binary executable programs. We demonstrate the versatility of this approach by using our representations to solve two semantically different binary analysis tasks -- functional algorithm classification and vulnerability discovery. We compare the proposed approach to our own strong baseline as well as published results, and demonstrate improvement over state-of-the-art methods for both tasks. 
We evaluated Bin2vec on 49191 binaries for the functional algorithm classification task, and on 30 different CWE-IDs including at least 100 CVE entries each for the vulnerability discovery task. We set a new state-of-the-art result by reducing the classification error by 40\% compared to the source-code based inst2vec approach, while working on binary code. For almost every vulnerability class in our dataset, our prediction accuracy is over 80\% (and over 90\% in multiple classes).

\end{abstract}

\section{Introduction}
For many security problems, researchers are relying on binary code analysis, as they need to inspect binary executable program files without access to any source code. 
This is often needed when analyzing commercial code that is protected by intellectual property and its source code is not available, but can be also useful in other scenarios. 
Those include dealing with unsupported or legacy executables, where the information about the exact version of the source code is lost, or even the original source code itself may be lost.
Additionally, it is frequently used for testing in order to improve the security of the system, like in black-box penetration testing when the goal is to check the binary for any weaknesses or vulnerabilities that can potentially be abused. 
And finally, it is an important part of investigating how hard recovering key parts of the algorithm is, \textit{e.g.,} for the sake of preventing intellectual property theft.

The translation process of going from source code to binary executable programs also called compilation, is a lossy process in the sense that only basic low-level instructions and data representations understood by the target CPU are preserved. Because of this, it is impossible in the general case to reconstruct the original source code from compiled binary code. The task is even more complicated with commercial binaries as they are often stripped. Stripping of the binary removes any debug information and its symbol tables, which contain semantics of variables in the program. When dealing with stripped binaries, even reconstructing function entry points can be challenging. 

With a constantly growing number of computing devices in consumer, commercial, industrial and infrastructure applications, as well as with the growing complexity of software applications, the scope of binary code analysis becomes increasingly large. Fast, automated analysis would allow preventing the spreading of bugs and vulnerabilities in all those complex software systems through shared and reused code. 

Analyzing binary executable code is difficult because of two related challenges  - the size of binary executable programs and the absence of high-level semantic structure in binary code. Indeed, when dealing with a compiled executable, a security engineer is often looking at a file containing up to megabytes of binary code. A precise analysis of such files with existing tools requires large amounts of computational power, and it is particularly difficult or even impossible to do manually. Instead, state-of-the-art tools often rely on a combination of formal models and heuristics to reason about binary programs. Replacing these heuristics with more advanced statistical learning and machine learning models has a high potential for improving performance while keeping the analysis fast. 

In recent years we have seen a big surge in applications of machine learning (ML) to the field of security, where researchers routinely turn to ML algorithms for smarter automated solutions.  For example, due to rapidly evolving modifications of malware, ML algorithms are frequently applied to malware detection problems. Similarly, ML algorithms allow detecting and reacting to network attacks faster. 

Having ML algorithms operate on binary executable programs is a promising direction to bridge the large semantic gap between human abstractions and machine code representations, and to recover high-level semantics which was lost during compilation. Using ML requires obtaining a good, vectorized representation of the data. In the field of security, this problem is usually solved by hand-selecting useful features and feeding those into an ML algorithm for a prediction or a score. Approaches range from defining code complexity metrics and legacy metrics~\citep{DBLP:conf/icse/TheisenHMMW15}, to using a sequence of system calls~\citep{DBLP:conf/codaspy/GriecoGURFM16} and many more. Besides being non-trivial and laborious, hand-selecting features raises other issues as well.
First, for every task researchers come up with a new set of features. For example, what indicates memory safety violations is unlikely to also signal race conditions. 
Additionally, some features get outdated and will need to be replaced with future versions of the programming language, compiler, operating system or computer architecture. 

The state-of-the-art in machine learning, however, no longer relies on hand-designed features. Instead, researchers use learned features, or what is called \emph{distributed representations}.
These are high-dimensional vectors, modeling some of the desired properties of the data. 
The famous word2vec model~\citep{DBLP:journals/corr/abs-1301-3781,DBLP:conf/nips/MikolovSCCD13}, for example, is representing words in a high-dimensional space, such that similar words are clustered together. 
This property of word2vec has made it a long-time go-to model for representing words in a number of natural language processing tasks. 
We can take another example from computer vision, where it was discovered that outputs of particular layers of VGG network~\citep{DBLP:journals/corr/SimonyanZ14a} are useful for a range of new tasks.

We see an important argument for trying to learn distributed representations - a good representation can be used for new tasks without significant modifications.
Unfortunately, some types of data are more challenging to obtain such a representation for, than others.
For instance, finding methods for representing longer sentences or paragraphs is still an ongoing effort in natural language processing~\citep{DBLP:conf/nips/ZhangSWGHC17,DBLP:conf/iclr/LinFSYXZB17}. 
Representing graphs and incorporating structure and topology into distributed representations is not fully solved either.
Binary executable programs are a ``hard'' case for representing as they have traits of both longer texts and structured, graph-like data, with important properties of binaries best represented as control or data flow graphs. 

Distributed representations for compiled C/C++ binaries -- the kind that engineers in the security field deal with the most -- have not received much attention, and with this work, we hope to start filling that gap. In fact, current approaches leveraging deep learning models to reason about binary code focus on code clone detection, and therefore, their application to algorithm classification and vulnerability detection is limited to syntactically similar patterns. In contrast, our approach aims to generalize code semantics  based on new insights by introducing a graph embedding model which encompasses notions of local control-flow and data-flow in a novel way. We propose a graph-based representation for binary programs, that when used with a Graph Convolutional Network (GCN)~\citep{DBLP:conf/iclr/KipfW17}, captures semantic properties of the program.


Our main contributions are: (i) To the best of our knowledge we are the first to suggest a distributed representation learning model approach for binary executable programs that is demonstrated to work for different downstream tasks;(ii)
To this end, we present a deep learning model for modelling binary executable programs’ structure, computations, and learning their representations; (iii) To prove the concept that distributed representations for binary executable programs can be applied to downstream programs analysis tasks, we evaluate our approach on two distinct problems - functional algorithm classification (\textit{i.e.,} the task of recognizing functional aspects of algorithmic properties, as opposed to their syntactic aspects) and vulnerability discovery across multiple vulnerability classes, and show improvement over current state-of-the-art approaches on both.  

\section{Related Work}
Many tasks that rely on the analysis of binary executables are frequently approached by rule-based systems and manually defined heuristics~\citep{DBLP:conf/securecomm/AaferDY13,DBLP:conf/iceis/SantosPDB09,DBLP:journals/di/KarbabDDM18,DBLP:conf/sp/YamaguchiGAR14,DBLP:conf/ssiri/RawatM12,DBLP:conf/sp/ChaARB12}.
Machine learning has a proven reputation for boosting performance compared to heuristics and there has been a lot of interest in applications of machine learning to security tasks.
We briefly discuss previous work in binary program analysis that relies on machine learning.
We structure the literature based on the types of features extracted and by the type of the embedding model applied. 
 
\subsection*{Hand Designed Features}
Designing and extracting features can be considered equivalent to manually crafting representations of binaries. 
We can classify such approaches based on which form of the compiled binary program was used to extract the features.

\paragraph{Code-based features}
The simplest approach to representing a binary is by extracting some numerical or textual features directly from the assembly code. This can be done by using n-grams of tokens, assembly instructions, lines of code, etc. N-grams are widely used in the literature for malware discovery and analysis \citep{DBLP:conf/ict/LiZYY19,DBLP:journals/tjs/LeeCSK18,DBLP:journals/ijcysa/KangYSM16}, as well as vulnerability discovery~\citep{DBLP:conf/icmla/PangXN15,DBLP:journals/jss/MurtazaKHB16}. 
Additionally, there have been efforts focusing on extracting relevant API calls or using traces of system calls to detect malware~\citep{DBLP:journals/infsof/WuWLZ16,DBLP:conf/ausai/KolosnjajiZWE16}.

\paragraph{Graph-based features}
Many solutions rely on extracting some numerical features of Abstract Syntax Trees (ASTs), Control Flow Graphs (CFGs) and/or Data Flow Graphs (DFGs). We combine these under models with graph-based features.
discovRE~\citep{DBLP:conf/ndss/EschweilerYG16}, among other features, uses closeness of control flow graphs to compute similarity between functions. 
Genius~\citep{DBLP:conf/ccs/FengZXCTY16} converts CFG into numeric feature vectors to perform cross-architecture bug search.
Yet other works have used quantitative data flow graph metrics to discover malware\citep{DBLP:conf/dimva/WuchnerOP15}.

\subsection*{Learned Features}
Besides manually crafting the representations it is also possible to employ neural models for that purpose.
This allows expressing and capturing more complicated relations of characteristics of code.
Here we can classify the approaches based on whether they use sequential neural networks or graph neural networks.  

\paragraph{Sequence embeddings}
The body of work on the naturalness of software~\citep{DBLP:journals/cacm/HindleBGS16,DBLP:conf/icse/RayHGTBD16,DBLP:journals/csur/AllamanisBDS18} has inspired researchers to try applying NLP models for security applications in general, and binary analysis in particular. 
Researchers have suggested serializing ASTs into text and using them with LSTMs for vulnerability discovery \citep{DBLP:conf/ccs/LinZLPX17}.
Some of previous vulnerability discovery efforts also use RNNs on lines of source code \citep{DBLP:conf/ndss/LiZXO0WDZ18}.
More recently, INNEREYE proposed to use LSTMs in a Siamese architecture for binary code similarity detection \citep{DBLP:conf/ndss/ZuoLYL0Z19}. 
The closest to our work is that of \cite{DBLP:conf/sp/DingFC19}, which is starting by constructing a graph that is enriched with selective callee expansion. The authors then sample random walks from this graph to generate sequences of instructions and train a paragraph-to-vector model on these sequences. This approach is similar in spirit to earlier graph embedding approaches, such as DeepWalk~\citep{DBLP:conf/kdd/PerozziAS14}, that were sampling random walks of nodes and using word embedding models on sequences of adjacent nodes for representation learning. However, today these approaches for graph embedding are no longer popular, as graph neural networks based on message passing and neighbourhood aggregation have been shown to perform much better.

\paragraph{Graph embeddings}
Graph embedding neural models are a popular choice for tackling binary code-related tasks because the construction of Control Flow or Data Flow Graphs is frequently an intuitive and well-understood first step in binary code analysis.  For instance, graph embedding models have successfully been used on top of Control Flow Graphs for tackling the task of code clone detection in source code and binary programs \citep{DBLP:conf/kbse/WhiteTVP16,DBLP:conf/ccs/XuLFYSS17,DBLP:conf/icml/LiGDVK19,DBLP:conf/nips/ZhouLSD019}. From these, Gemini \citep{DBLP:conf/ccs/XuLFYSS17} uses a Siamese architecture on top of a graph embedding model for binary code clone detection task. The graphs they use - attributed control flow graphs, or ACFGs, - are CFG graphs that are enriched with a few manually defined features. In our work, instead of enhancing the basic blocks in CFG with a few attributes, we suggest enriching them by expanding the computations in each basic block into a computational tree, and rely on the fact that the graph embedding model will be able to capture attributes like the number of instructions if necessary. Graph Matching Networks (GMNs), \citep{DBLP:conf/icml/LiGDVK19} on the other hand, are based on the idea that instead of computing an embedding and then either using a distance function, or a Siamese network for the comparison, it might be beneficial to directly compare two graphs. So, as opposed to Gemini, where representations for known vulnerable or benign programs were pre-computed, GMN needs to compute similarity for every pair of programs individually starting from scratch. They demonstrate that this approach has better performance compared to Siamese architectures, but it is clearly slower, and more importantly for us - it does not produce program embeddings.

Other research using graph structure of binary programs include using Conditional Random Fields on an enhanced Control Flow Graph for attempting to recover the debug information of the binary program~\citep{DBLP:conf/ccs/HeITRV18}.

\begin{figure*}
\begin{subfigure}{.3\textwidth}
\centering
\includegraphics[width=\textwidth]{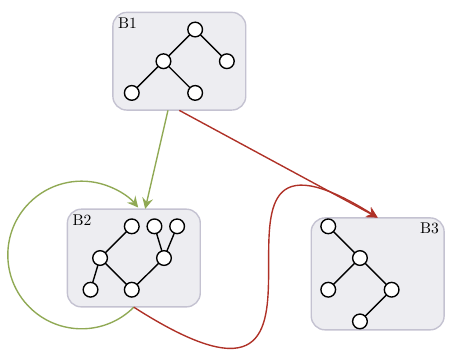}
\caption{}
\label{fig:a}
\end{subfigure}
\begin{subfigure}{.38\textwidth}
\centering
\includegraphics[width=\textwidth]{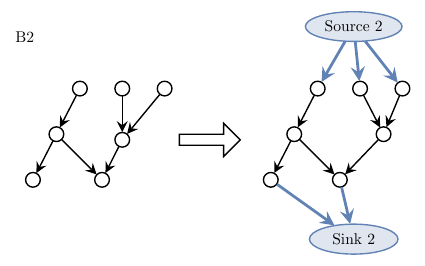}
\caption{}
\label{fig:b}
\end{subfigure}
\begin{subfigure}{.23\textwidth}
\centering
\includegraphics[width=\textwidth]{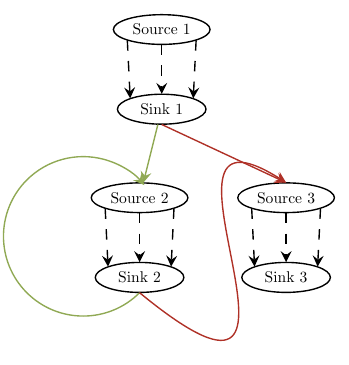}
\caption{}
\label{fig:c}
\end{subfigure}
\caption{This figure illustrates how the program graph is constructed from the CFG on a toy example with three basic blocks. Fig (a) demonstrates, that for each basic block independently we construct a computational tree. On fig (b) we can see, that each tree is completed with two nodes - a source and a sink, this way turning it into a directed acyclic graph (DAG). On fig (c) we demonstrate how all DAGs are connected following the same topology as CFG had initially.}
\label{fig:program_graphs}
\end{figure*}

\begin{figure*}
\centering
\includegraphics[width=0.9\textwidth]{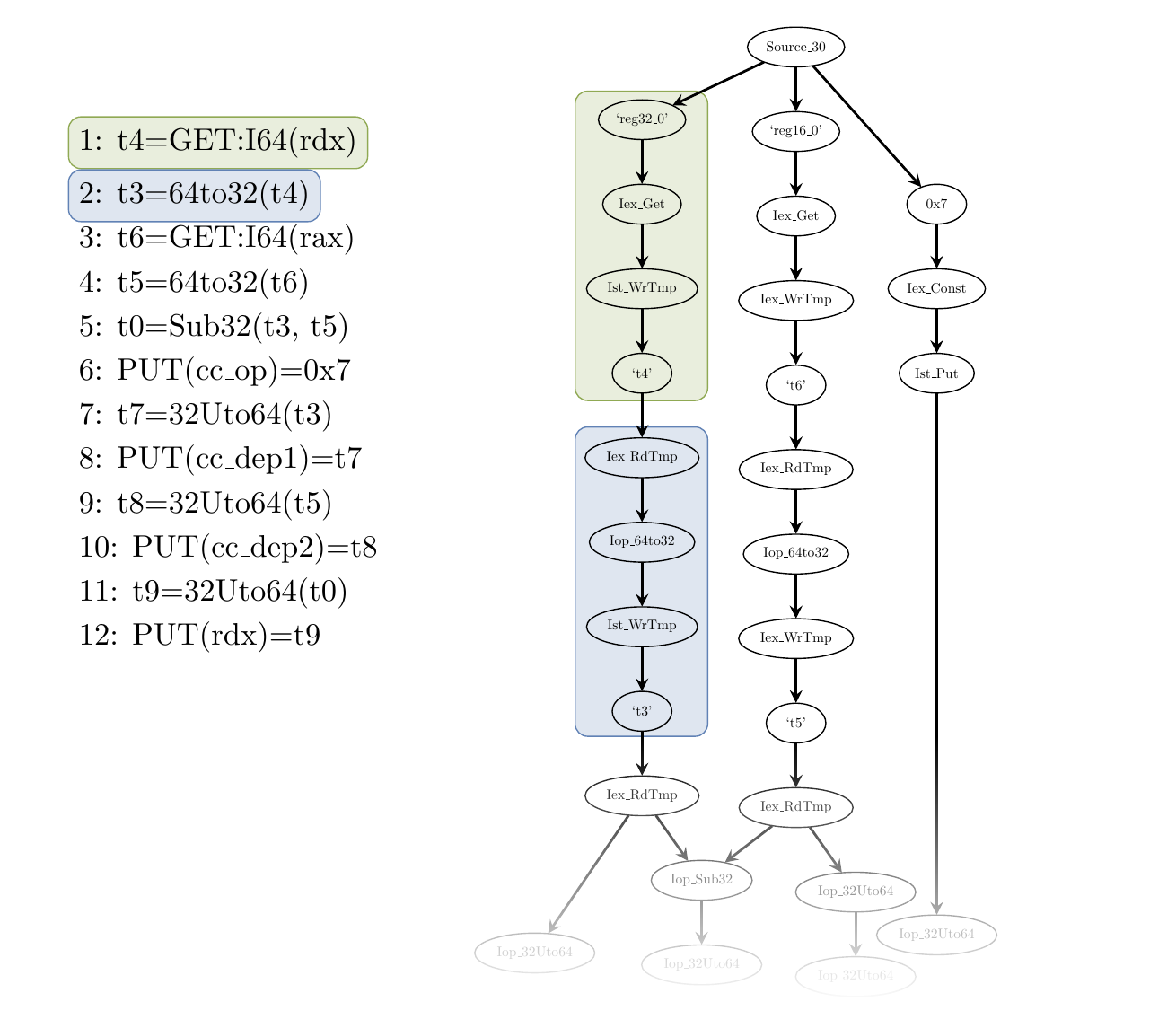}
\caption{An example of the program graph. Parts of the graph are highlighted in the same color, as instructions on lines 1 and 2, to demonstrate where those instructions were mapped to in the graph.}
\label{fig:inst_to_tree}
\end{figure*}

\begin{figure*}
\centering
\includegraphics[width=0.9\textwidth]{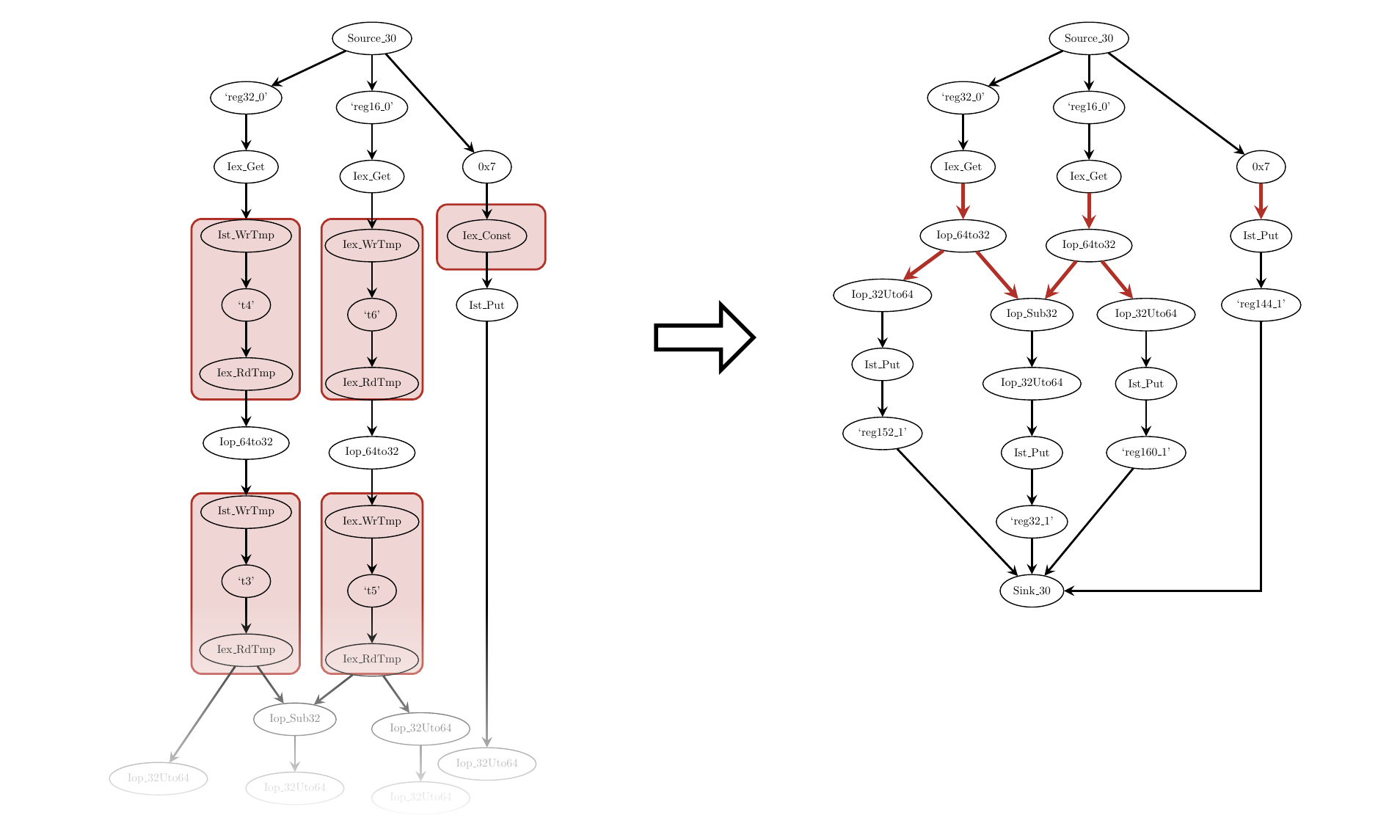}
\caption{Here we show how redundant instructions were removed to contract the graph. On the left the graph is shown before the contraction, and on the right it is demonstrated after the contraction.}
\label{fig:redundant_inst}
\end{figure*}

\section{Model}
We start by converting the binary executable to a program graph that is designed to allow mathematically capturing the semantics of the computations in the program. 
Next, we use a graph convolutional neural network to learn a distributed representation of the graph. 
Below we describe the process of constructing the program graph, followed by a brief introduction to how graph convolutional neural networks work. We also describe the baseline model that we use for evaluation and comparison, alongside  previous existing approaches.

\subsection{Program Graphs}\label{sec:programgraphs}
We start by disassembling the binary program and constructing a control flow graph (CFG). We use static inter-procedural CFGs, which we construct using the angr library \citep{DBLP:conf/sp/Shoshitaishvili16}.

The fact that each basic block in CFG is executed linearly allows us to unfold the instructions within each basic block and represent them as a directed, computational tree, similar to an Abstract Syntax Tree (AST). The result of this process is schematically depicted in Figure~\ref{fig:a}. 

Within each basic block, computations do not necessarily all depend on each other. 
There may be chunks of code that can be reordered inside the basic block without affecting the final result.
In this case the approach described so far yields a forest of computations. 
To connect the trees in the forest we add \textit{Source} and \textit{Sink} nodes at the beginning and at the end of each basic block as a parent, or correspondingly a child, for all the trees generated from that basic block, which is demonstrated in Figure~\ref{fig:b}. The resulting graphs are then connected following the same topology that basic blocks originally had in the CFG, as shown in Figure~\ref{fig:c}.

We construct the above-mentioned computational trees from VEX intermediate representation (IR) of the binary. Figures~\ref{fig:inst_to_tree}~and~\ref{fig:redundant_inst} provide demonstration of the process.

Every node of the resulting tree is thus labelled with a constant, a register, a temporary or an operation. 
The edges of the tree are directed from the argument to the instruction. 
Within each basic block we reuse nodes that correspond to addresses, temporaries, constants and registers to tie together related computations.
VEX IR provides Static Single Assignment form (SSA). This means that each assembly instruction in a basic block is lifted and ``spilled'' into multiple IR statements operating on temporary variables that are each used only once (the goal being to make all side effects of an instruction explicit). 
However, VEX does not track instances of different \emph{definitions} and \emph{uses} of the same register across instructions within the basic block, which we implemented to ensure we do not introduce fake data-dependence edges. In our implementation, if an instruction overrides or redefines the content of a register, its subscript is incremented. For example, for the \texttt{eax} register, we start from \texttt{eax\_0} and increment it to \texttt{eax\_1}. This is necessary so that we do not reuse the same node for \texttt{eax\_0} and \texttt{eax\_1}.

As a last step, we remove redundant edges and nodes, particularly, the 
$\texttt{Iex\_Const}$
node that follows every constant, and chains of 
$\texttt{Iex\_WrtTmp}\rightarrow$
`t\%'
$\rightarrow\texttt{Iex\_RdTmp}$\footnote{In VEX $\texttt{Iex\_Const}$ represents a constant value, $\texttt{Iex\_WrtTmp}$ a write operation (to a temporary variable), and $\texttt{Iex\_RdTmp}$ a read operation (from a temporary variable)}. This is demonstrated in Figure~\ref{fig:redundant_inst}. 

After the graph construction is complete, we remove SSA indices for temporary variables and registers to reduce the number of distinct labels. 

From the labels of the nodes we construct a ``feature matrix'' of the graph, which is a matrix of size $n\times d$, where $n$ is the number of nodes in the graph, and $d$ is the number of all distinct labels seen in the entire dataset. Thus, every node has one row in the feature matrix associated with it. We choose a random fixed ordering of all labels, and then for a given node, to convert its label into its feature row we assign all positions of the row to zero with the exception of the position that corresponds to the label of the node in our fixed ordering of labels. This representation is known as a one-hot representation. We will further refer to the feature matrix as $X$. Note that we use words ``feature'' or ``features'', ``embeddings'' and ``representation'' interchangeably.

\subsection{Graph Convolutional Networks}\label{sec:gcn}

\begin{figure*}
\centering
\includegraphics[width=0.95\textwidth]{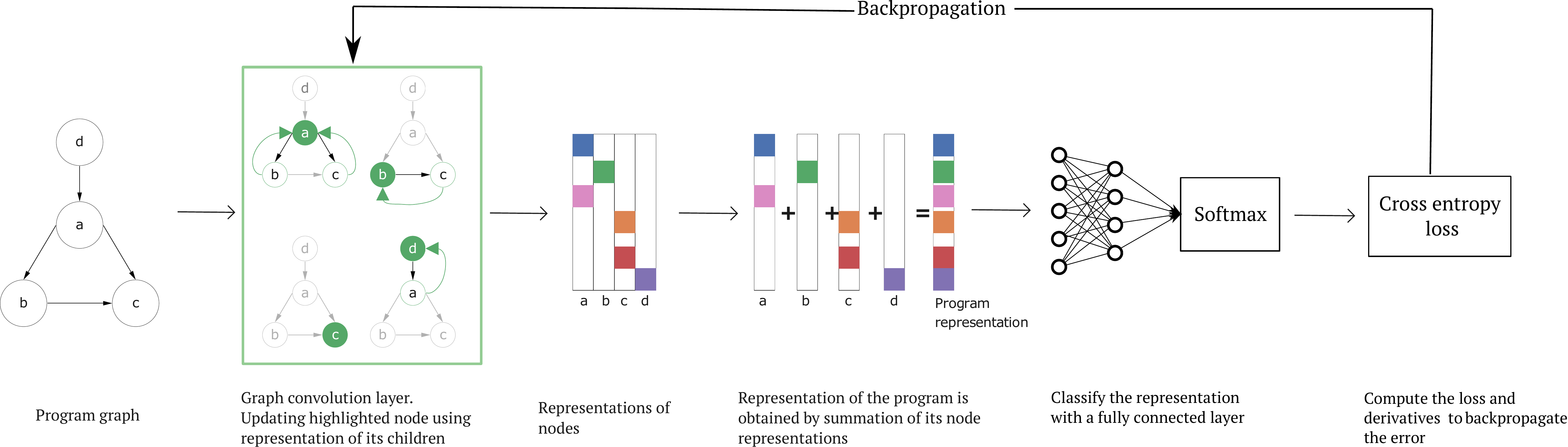}
\caption{Schematic depiction of obtaining a program representation with a single-layer GCN model}
\centering
\label{fig:gcn}
\end{figure*}
The model we used for learning representations is a Graph Convolutional Neural Network (GCN) \citep{DBLP:conf/iclr/KipfW17}. Graph neural embeddings is a fast developing field, and some alternative graph representation learning models include GraphSAGE \citep{DBLP:conf/nips/HamiltonYL17} or Gated Graph Neural Networks \citep{DBLP:journals/corr/LiTBZ15}, as well as a number of others. In the literature GCNs consistently perform on par or better than more recent variants of graph neural networks \citep{DBLP:conf/cvpr/MontiBMRSB17,DBLP:conf/aaai/LiuCLZLSQ19,DBLP:journals/corr/abs-1710-10903,DBLP:conf/iclr/ChenMX18}, while being simpler and oftentimes, faster. We chose GCN because it provides a good trade-off between simplicity, performance, and speed. The latter is important due to the low-level nature of the binary code; it is reasonable to expect the program graphs to grow quite large, which forces us to favour a model with weight updates that can be efficiently computed in batches.

GCN consists of a few stacked graph convolutional layers.
A graph convolutional layer is applied simultaneously to all nodes in the graph. For each node, it averages the features of that node with features of its neighbours. Features of different nodes are scaled differently in the process of averaging and these weights are learned, i.e. they are the parameters of the graph convolutional layer. After the averaging, each node is assigned the resulting vector as its new feature vector and we proceed to either apply a different graph convolutional layer, or compute the loss and perform backpropagation to update the parameters.

Formally, this process of computing new feature vectors, known as forward pass or propagation, for $(l+1)$-st graph convolutional layer can be described as follows:
\begin{equation}
    H^{(l+1)} = \text{ReLU}
    (D^{\frac{1}{2}}
    \Tilde{A}
    D^{\frac{1}{2}}
    H^{l} W^{l})
\end{equation}
where $\Tilde{A}$ is the adjacency matrix of the graph with added self-loops, $D$ is its diagonal out-degree matrix, $\text{ReLU}(x) = max(0, x)$ is the non-linearity or activation function, $H^{(l)}$ is the result of propagation through previous layer, $H^{(0)}$ being $X$, and $W^{l}$ is a layer-specific trainable weight matrix.

Since one graph convolutional layer averages representations of the immediate neighborhood of the node, after performing $k$ graph convolutions we incorporate the information from $k$-th neighborhood of the node.

From our description, it follows that after the forward pass, the graph convolutional network outputs features for each node in the graph. We will refer to this new feature matrix as $Z$. Note that $Z$ still has $n$ rows - one row per node, but it can have a different number of columns.  

To get the representation of the entire graph, we can aggregate the features of all nodes in the graph. Here it is possible to use any aggregation function - summation, averaging, or even a neural attention mechanism, but in our experiments we went for a simple sum aggregate.
A schematic illustration of this entire process is available in Figure~\ref{fig:gcn}.

The aggregated representation is used with a two-layer perceptron, and passed through a softmax which is defined like $\text{softmax}(x_i) = \frac{exp(x_i)}{\sum_{i}{exp(x_i)}}$, for the final classification.

We frame our tasks as classification and use cross-entropy error as the objective function for the optimization.We cover our procedure for selecting hyperparameters for GCN model in more detail in Section~\ref{sec:task1-exp-setup}.

\subsection{Baselines}\label{sec:baseline}
We wanted to compare our proposed representation to another task-independent representation, in particular, to one that used code-based features or embeddings.
We experimented with Long Short Term Memory (LSTM) neural networks and Support Vector Machine (SVM) classifiers for that purpose. We interpreted instructions as words, and a sequence of instructions as a sentence, following a number of similar approaches in the field, e.g. \cite{DBLP:conf/ndss/ZuoLYL0Z19}. We experimented using both SVM and LSTM with the assembly instructions directly, as well as with the code lifted to VEX IR. From our experiments, an SVM classifier with a Gaussian kernel and bag-of-words representation of VEX IR gave us the best performance, so that is the setup we chose as a baseline.
Each line of IR is tokenized to be a single ``word''. 
Vocabulary for the bag-of-words was obtained from the training part of the dataset. 
We used frequency thresholding to remove infrequent entries and reduce data sparsity. 
Those frequencies were empirically found on the validation part of the dataset.

\begin{table*}
\centering
\caption{An example prompt for programming competition problems and their corresponding problem numbers and names. The example is taken from ACM Timus \url{http://acm.timus.ru/}}
\begin{tabular}{ll}
\toprule
\multicolumn{1}{c}{\textbf{Prompt}} & \multicolumn{1}{c}{\textbf{Problem \#}} \\
\midrule
\begin{tabular}{p{0.70\textwidth}}You have a number of stones with known weights $w_1, \ldots w_n$. Write a program that will rearrange the stones into two piles such that weight difference between the piles is minimal.\end{tabular} & 1005. Stone Pile \\
\hline
\end{tabular}
\label{tab:timus}
\end{table*}

\section{Task Description}
We evaluate the performance of our proposed representations on two independent tasks. 
In the first, we test the proposed representations for functional algorithm classification in binary executable programs through classifying coding challenges.  
In our second task, we want to demonstrate the performance of learned representations on a common security problem -- discovery of vulnerable compiled C/C++ files.
The two tasks are semantically different and we demonstrate in the later sections that both can be successfully tackled with representations constructed and learned in the same way. 
 
\subsection{Task 1: Functional Algorithm Classification}
Algorithm classification is crucial for semantic analysis of code. We qualify it as ``functional'' by opposition to ``syntactic'', i.e., we aim to capture the semantics of functional properties of algorithms.
It can be used for creating assisting tools for security researchers to understand and analyze binary programs, or discover inefficient or buggy algorithms, etc.

In this task, we are looking at real-world programs submitted by students to solve programming competition problems. We chose such a dataset because the programs in it, being written by different students, naturally encompass more implementation variability than it would be possible to get by using, for instance, standard library implementations. 
Our goal is to classify solutions by the problem prompts that the solution was written for. 

We present a typical example of programming competition problem prompt in Table~\ref{tab:timus}. 
Provided example is for illustrative purposes only, as it is taken from ACM Timus (\url{http://acm.timus.ru}) and is not part of our dataset\footnote{The dataset we used was collected as part of previous work for which the problem prompts are not released with the data}.

From our definition and the dataset, it follows that we define the equivalence of two programs as them solving the exact same problem. 
Hence, in this task, we test the \emph{ability of the model to capture the higher-level semantic} similarity, and to take into account program behaviour, functionality and complexity, while \emph{ignoring syntactic differences} wherever possible. 

\subsection{Task 2: Vulnerability Discovery}
Software contains bugs, which in the worst case can lead to weaknesses that leave vulnerable systems open to attacks. 
Such security bugs, or vulnerabilities, are classified in a formal list of software weaknesses - Common Weakness Enumeration (CWE). 
Vulnerability discovery is the process of finding parts of vulnerable code that may allow attackers to perform unauthorized actions. It is an important problem for computer security. The typical target of vulnerability discovery is programming mistakes accidentally introduced in \emph{benign commodity programs} by their authors. Our work excludes software specifically crafted to behave in a malicious way, and focuses on benign programs. Due to the large variability among vulnerabilities, increasingly large sizes of software and increasing costs of testing it, the problem of vulnerability discovery is not solved. 

Most vulnerability discovery techniques rely on dynamic analysis for program exploration, the most common one being fuzzing~\citep{afl}. Such models offer a high level of precision, at the cost of shallow program coverage: only a subset of execution traces for a given program (along with a set of input test cases) can be observed in finite time, leaving large parts of the program unexplored.
On the other hand, static analysis provides better program coverage at the cost of lower precision. 
In addition to these challenges come a range of fundamental problems in program analysis related to undecidability (\textit{e.g.,} the halting problem, \textit{i.e.,} ``Does the program terminate on all inputs?'') and implementation.
These issues emerge because vulnerabilities may span very small or very large chunks of code and involve a range of different programmatic constructs. This raises the question - at what level of granularity in the program should we inspect them for vulnerabilities or report to security researchers.
In this work, we are concerned with the question of learning representations for the entire binary program that will help to discover vulnerabilities statically, while leaving the questions of handling large volumes of source code and working on variable levels of granularity for future work. Our work builds on standard binary-level techniques for control-flow recovery (i.e., the reconstruction of a CFG), which is a well-studied problem where state-of-the-arts models perform well with high accuracy and scalability~\citep{victor}.

\section{Datasets and Experimental Setup}
Our first dataset, introduced by \citep{DBLP:conf/aaai/MouLZWJ16}, consists of 104 online judge competition problems and 500 C or C++ solutions for each problem submitted by students. 
We only kept the files that could be successfully compiled on a Debian operating system, using gcc8.3, without any optimization flags. This left us with 49191 binary executable files, each belonging to one of 104 potential classes. Each class in this dataset corresponds to a different problem prompt and our goal is to classify the solutions according to their corresponding problem prompts. 

The second dataset we used is the Juliet C/C++ test suite\citep{DBLP:journals/computer/BolandB12}. 
This is a synthetically generated dataset, that was created to facilitate research of vulnerability scanners and enable benchmarking. 
The files in the dataset are grouped by their vulnerability type -- CWE-ID. 
Each file consists of a minimal example to recreate the vulnerability and/or its fixed version. 
Juliet test suite has $OMITGOOD$ and $OMITBAD$ macros, surrounding vulnerable and non-vulnerable functions correspondingly.
We compiled the dataset twice - once with each macro, to generate binary executable files that contain vulnerabilities and those that do not. 
The dataset contains 90 different CWE-IDs.
However, some of them consist of Windows-only examples, that we omitted.
Note that even though our approach is not platform-specific, in this work we limit our experimentation to Linux only.

Most CWE-IDs had too few examples to train a classifier and/or to report any meaningful statistics on.\footnote{In the future, we consider combining some CWEs into their umbrella categories, for example following the classification by Research Concepts:\\ \mbox{\url{https://cwe.mitre.org/data/definitions/1000.html}}}
Thus, we also omitted any CWE-ID that had less than 100 files in its testing set after 70:15:15 for training:validation:test split, because for those cases the reported evaluation metric would be too noisy.
As a result, we experimented on vulnerabilities belonging to one of 30 different CWE-IDs, presented in Table~\ref{tab:juliet-cwe-counts}.
We trained a separate classifier for every individual CWE-ID, which was required because files associated with each CWE-ID may or may not contain other vulnerability types.

We trained the neural network model with early stopping, where the number of training epochs was found on the validation set.

\subsection{Task 1. Experimental Setup}\label{sec:task1-exp-setup}
For experiments in the functional algorithm classification task, we randomly split all the binaries in the first dataset into train:test:validation sets with ratios 70:15:15. 
We use the training set for training and extracting some additional helper structures, such as vocabulary for the bag of words models and counting frequencies for thresholding in neural network models. 
We use the validation set for model selection and finding the best threshold values. After finding the best model, we evaluate its performance on the testing set. 
The experiments are cross-validated and averaged over 5 random runs.

For SVMs, in the model selection phase, we perform a grid-search over the penalty parameter C and pick a value for the vocabulary threshold to remove any entry that does not have a substantial presence in the training set to be useful for learning. 
After the trimming our vocabulary contains about 10-11K entries (the exact number changes from one random run to another).

For GCN-based representation, we follow similar logic and use the training set to find and remove infrequent node labels.
Here too the exact threshold is decided via experimentation on the validation set. 
On average, we keep about 7-8K different node labels. 
Very infrequent terms are replaced with a placeholder $UNK$, or $CONST$ if it is a hexadecimal.

\begin{figure*}
\centering
 \includegraphics[width=0.9\textwidth]{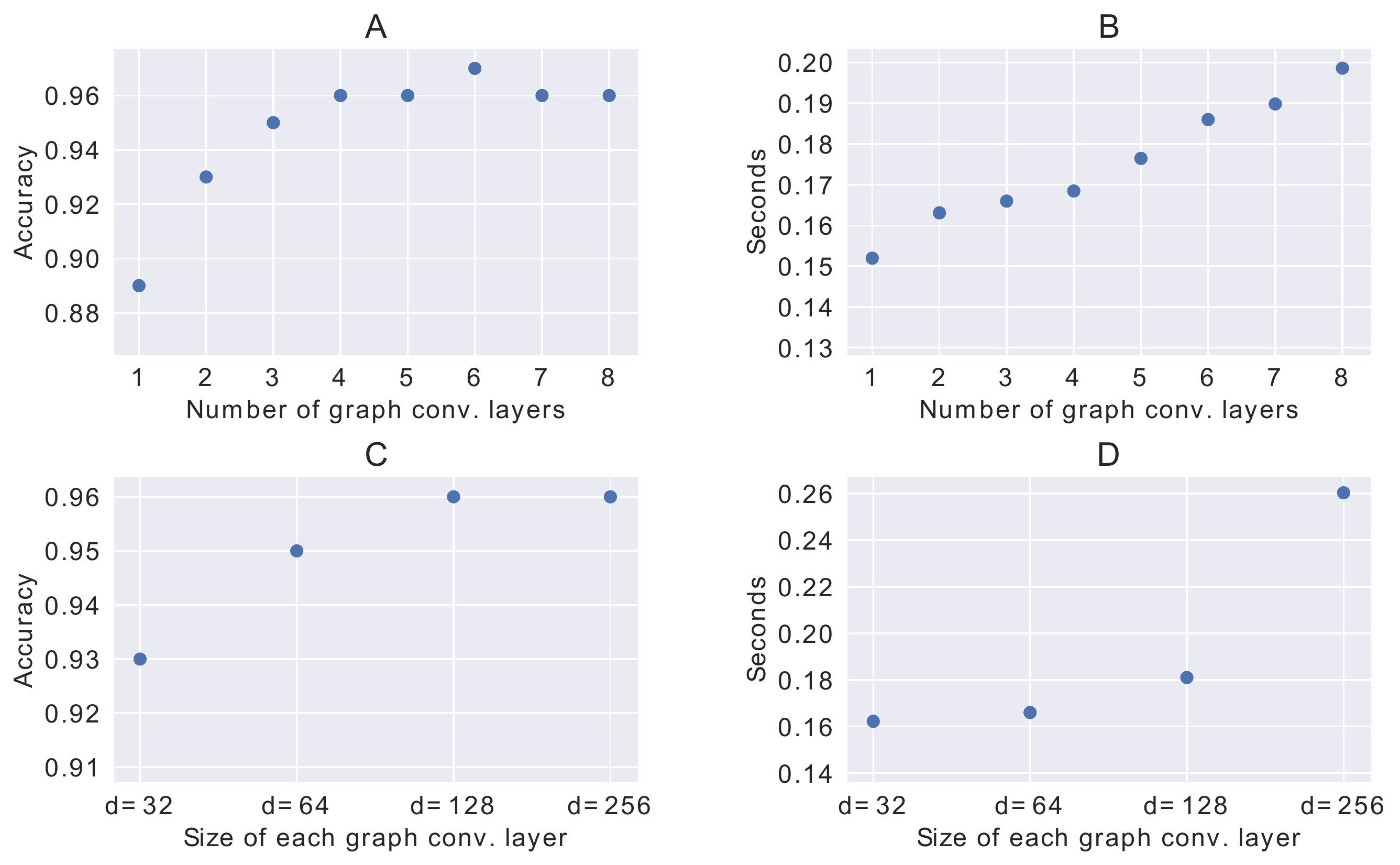}
\caption{Comparison of GCN models with different numbers of layers (A), and different sizes of each layer(C).\textbf{(A)} demonstrates the accuracy of GCN models with different numbers of layers - from 1 to 8 - on the validation set of Task 1; the size of each layer is 64. \textbf{(B)} shows the time in seconds that models from (A) take to train on 100 samples. \textbf{(C)} demonstrates the accuracy of GCN models with 3 layers, but different sizes of layers - from 32 to 256. \textbf{(D)} shows the time in seconds that models from (C) take to train on 100 samples.}
\label{fig:hyperparams}
\end{figure*}

We pick hyperparameters of the GCN model by their performance on the validation set. Figure~\ref{fig:hyperparams} demonstrates the influence of the \textit{depth} (number of graph convolution layers) and \textit{width} (size of each graph convolution layer) on the performance of the model for Task 1. Figure~\ref{fig:hyperparams}\textbf{(A)} shows the peformance of models with depths from 1 to 8 layers, while the dimensionality of every layer is set to 64. As it can be seen, increasing the depth of the model up until 4 layers improves performance, however additional layers after that do not always improve performance. Figure~\ref{fig:hyperparams}\textbf{(C)} compares performances of four models where each model has the same number of layers (3), but different sizes of layers - 32, 64, 128 or 256. From here we see, that increasing the size of the layers from 64 to 128 provides a moderate improvement, but increasing the size further does not affect the performance.  Figures~\ref{fig:hyperparams}\textbf{(B)} and \textbf{(D)} show the duration of training in seconds of each of the discussed above models on 100 examples. Based on these general findings we perform some additional experimentation and deploy a GCN with 3 layers, that has 128 dimensions in its first two layers, and 64 dimensions in its last layer. 

\begin{table*}
\centering
\caption{In this table we provide total counts of binary executables for each of the CWE-IDs we studied in the Juliet Test Suite.}
\label{tab:juliet-cwe-counts}
\begin{tabular}{ll@{\hskip 0.10\textwidth}ll
    @{\hskip 0.10\textwidth}ll}
\toprule
CWE-ID & \# examples & CWE-ID & \# examples & CWE-ID & \# examples \\
\midrule
CWE121 & 9486  &  CWE197 & 2664 &  CWE476 & 888  \\
CWE122 & 11946 &  CWE23  & 2960 &  CWE563 & 1116 \\
CWE124 & 3612  &  CWE36  & 2960 &  CWE590 & 6954 \\
CWE126 & 2639  &  CWE369 & 2736 &  CWE606 & 760  \\
CWE127 & 3612  &  CWE400 & 2280 &  CWE617 & 918  \\
CWE134 & 3800  &  CWE401 & 4176 &  CWE680 & 1776 \\
CWE190 & 12093 &  CWE415 & 2588 &  CWE690 & 2368 \\
CWE191 & 9048  &  CWE416 & 888  &  CWE758 & 1046 \\
CWE194 & 3552  &  CWE427 & 740  &  CWE761 & 888  \\
CWE195 & 3552  &  CWE457 & 2104 &  CWE762 & 6429 \\
\bottomrule
\end{tabular}
\end{table*} 

\subsection{Task 2. Experimental Setup}
In the vulnerability discovery experiments, we train a separate classifier for each of 30 different CWE-IDs. 
Note, that for each CWE-ID classifier in its training and testing we only include the binaries that are specifically marked as good or bad with regard to that CWE-ID. 
For every CWE-ID, we split its corresponding binaries into train:validation:test with ratios 70:15:15, and report results averaged over 5 random runs. 
We use training sets for training the models and validation sets for grid search of the penalty parameter C in SVMs. We report the performance of the best model measured on testing sets.
Here we reuse some statistics obtained on the first dataset, in particular, we reuse frequency thresholds and bag-of-words vocabularies.
We need to train a separate classifier for each CWE-ID, 30 SVM classifiers and 30 NN classifiers in total, which would lead to a huge search space at the phase of the model selection.

We are not aware of related work on vulnerability discovery that performs their evaluation on Juliet Test Suite. Thus, to give the readers a better understanding of how our proposed model would fare compared to other existing approaches, we performed an additional experiment using Asm2Vec model\citep{DBLP:conf/sp/DingFC19} on the Juliet Test Suite.
Asm2Vec is a clone search engine that relies on vector representations of assembly functions. In the original paper the authors suggested its usefulness as a vulnerability detection tool which allows finding duplicates of known vulnerable functions. We tried replicating that scenario as faithfully as possible, by training Asm2Vec\footnote{We used implementation available here: https://github.com/Lancern/asm2vec} on Juliet Test Suite, and comparing resulting representations to differentiate between vulnerable and non-vulnerable instances. Since Asm2Vec poses the vulnerability detection as a retrieval problem, we follow their example in the paper and report Precision@15 metric. For each vulnerable function, we find 15 most similar functions to it according to cosine similarity and compute the percentage of vulnerable functions among them. It is worth noting that we are looking for similar functions among all vulnerable and non-vulnerable functions per CWE-ID.

We set most of the hyperparameters of Asm2Vec following the original paper, but finetune for dimensionality of the representation and learning rate. To find best values for those we use grid search in intervals \{50,100,150,200\} and \{0.05, 0.025, 0.01 \}  correspondingly. The final results that we report for Asm2Vec are computed on the testing set, and are the average of 5 random runs.



\begin{figure*}
\centering
 \includegraphics[width=0.9\textwidth]{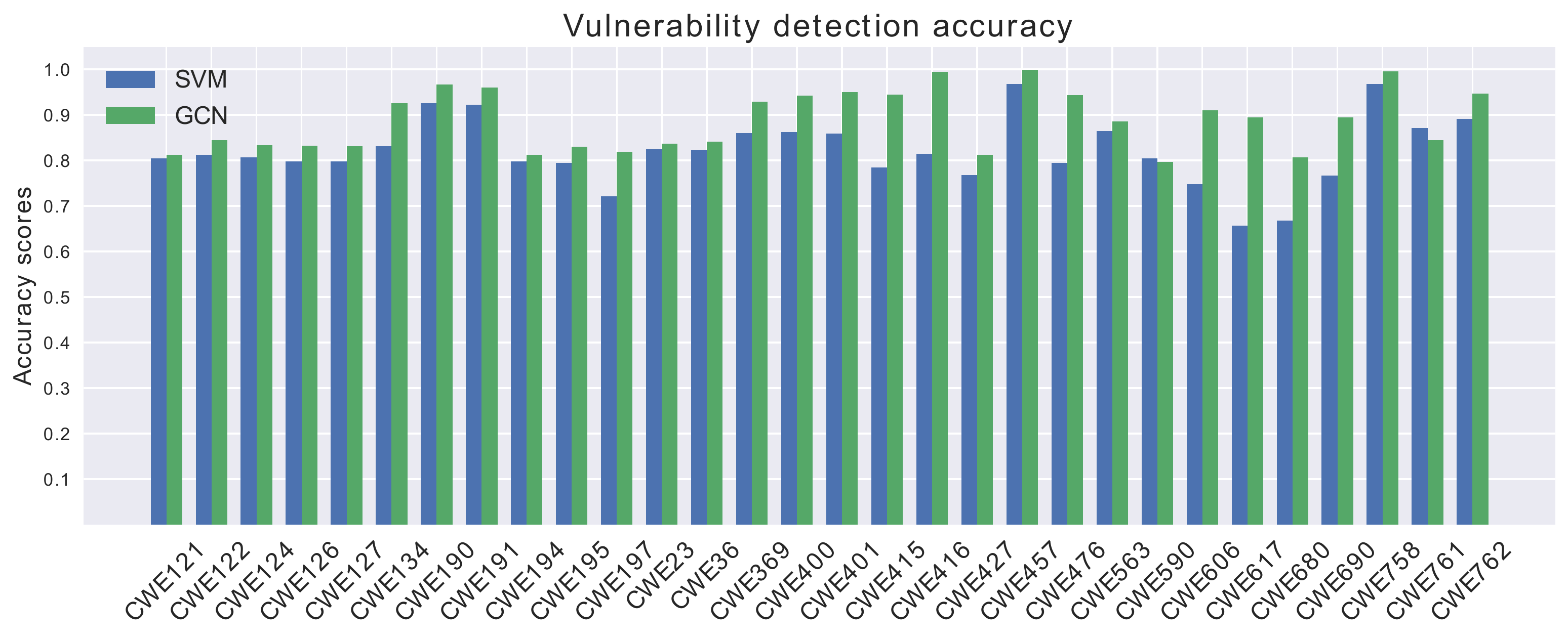}
\caption{Experimental results for vulnerability discovery on the Juliet test suite.}
\label{fig:julietresults}
\end{figure*}

\begin{table}
\centering
\caption{Accuracy obtained for the first task on the online judge problem classification.}
\label{tab:task1}
\begin{tabular}{lll}
\toprule
Model & Accuracy \\
\midrule
SVM on VEX IR & 0.93 \\
TBCNN~\citep{DBLP:conf/aaai/MouLZWJ16} & 0.94\\
inst2vec~\citep{DBLP:conf/nips/Ben-NunJH18} & 0.9483 \\
Ours & 0.97 \\
\bottomrule
\end{tabular}
\end{table}

\section{Evaluation and Results}
For evaluating performance in our experiments we used accuracy following previous work that we proceed to compare our results to.

\subsection{Task 1}
Table~\ref{tab:task1} contains quantitative evaluation of our representation for Task 1.
Our proposed representation outperforms our own SVM baseline, TBCNN model\citep{DBLP:conf/aaai/MouLZWJ16}, and current state-of-the-art for this task -  \textit{inst2vec}\citep{DBLP:conf/nips/Ben-NunJH18}. We manage to reduce the error by more than 40\%, thus setting a new state-of-the-art result. It should be additionally mentioned that both TBCNN and \textit{inst2vec} start from the C source code of the programs to make predictions, whereas our baseline SVM and our proposed model are only using compiled executable versions.

Highlighting a few important differences between our approach and \textit{inst2vec} helps better understanding some of the contributions of our approach. 
To construct the contextual flow graphs, the authors of \textit{inst2vec} compile the source code to LLVM IR, which contains richer semantic information than VEX IR that we use in this work. Because it is more high-level, LLVM IR is a difficult target for lifting from binary executable files.\footnote{More discussion on this topic is provided at Angr's FAQ page: \url{https://docs.angr.io/introductory-errata/faq}}. 

Another key difference is that instead of learning the representations of individual tokens and then combining the tokens into a program using a sequential model, we learn the representations of all the tokens in the program jointly, thus learning the representation of the entire program. The \textit{inst2vec}, on the other side, ignores the structural properties of the program at that step. 
Our results show that we can achieve better performance, despite \textit{inst2vec} starting from a semantically richer LLVM IR. We believe this indicates the importance of using the structural information at all stages of learning for obtaining good program embeddings.

\begin{table}[!ht]
   \caption{Asm2vec Precision@15 on Juliet Test Suite}\label{tab:asm2vec}
   \centering
   \begin{tabular}{ll@{\hskip 0.17\textwidth}ll
    @{\hskip 0.17\textwidth}ll}
   \toprule
     CWE-ID & P@15 & CWE-ID & P@15 & CWE-ID & P@15  \\ \midrule
     CWE121 & 0.72 & CWE197 & 0.71 & CWE476 & 0.73\\
     CWE122 & 0.61 & CWE23  & 0.68 & CWE563 & 0.63\\
     CWE124 & 0.63 & CWE36  & 0.68 & CWE590 & 0.64\\
     CWE126 & 0.69 & CWE369 & 0.54 & CWE606 & 0.56\\
     CWE127 & 0.66 & CWE400 & 0.56 & CWE617 & 0.64\\
     CWE134 & 0.66 & CWE401 & 0.50 & CWE680 & 0.75\\
     CWE190 & 0.53 & CWE415 & 0.66 & CWE690 & 0.55\\
     CWE191 & 0.59 & CWE416 & 0.77 & CWE758 & 0.72\\
     CWE194 & 0.71 & CWE427 & 0.47 & CWE761 & 0.55\\
     CWE195 & 0.72 & CWE457 & 0.69 & CWE762 & 0.58\\
     \bottomrule
   \end{tabular}
\end{table}

\subsection{Task 2}

Figure~\ref{fig:julietresults} contains the evaluation of our representation for Task 2.
Here, the classifier based on our proposed representation outperforms our SVM baseline in all cases except 2 -- CWE-ID590, Free of Memory not on the Heap, and CWE-ID761, Free of Pointer not at Start of Buffer. In both cases we are seeing less than 5\% difference in accuracy. On the other hand, our proposed representation demonstrates a significant gain in terms of performance. In the extreme case of CWE-617, Reachable Assertion, it outperforms the baseline by about 25\%, in many other cases the gain is from 10\% to 20\% of prediction accuracy. 


Table~\ref{tab:asm2vec} reports the results we obtained from running Asm2Vec on Juliet Test Suite. It is important to keep in mind that these numbers are not directly comparable to our results, as they correspond to two different metrics. Rather, this experiment demonstrates the complexity of the dataset and the capacity of Asm2Vec to capture vulnerabilities on it. While Bin2Vec achieves more than 80\% accuracy for all CWE-ID, Asm2Vec has Precision@15 equal to 0.5 or 0.6 in many cases, which means only about half of the retrieved similar functions were in fact vulnerable. Asm2Vec has highest Precision@15 of 0.77 for CWE-ID 416, Use After Free, which corresponds to about 1 in 4 retrieved functions being incorrectly labelled as vulnerable. For comparison, for the same vulnerability type Bin2Vec achieves near perfect performance.

Additionally, we can indirectly compare our results for the second task with those presented in two surveys that use Juliet Test Suite as a benchmark for evaluating commercial static analysis vulnerability discovery tools~\citep{DBLP:conf/csiirw/VelichetiFPRH14,DBLP:journals/infsof/Goseva-Popstojanova15}. It must be noted, that the commercial tools in those experiments probably did not use most of the programs for each CWE-ID as a training set. Additionally, the tools considered in those surveys are making their predictions based on source code and not binaries. Nevertheless, the comparison of the reported accuracies in those surveys with ours tells us that our proposed representation performs better for vulnerability discovery than static analysis commercial tools. For example, on CWE-IDs from 121 to 126, which are all memory buffer errors, \citep{DBLP:conf/csiirw/VelichetiFPRH14} report less than 60\% accuracy, whereas our model scores higher than 80\% for each of those CWE-IDs. For tools studied in \citep{DBLP:journals/infsof/Goseva-Popstojanova15}, our model consistently outperforms three out of four static analysis tools, and for the last one it outperforms it by a considerable margin in all cases but two. Those two are CWE-ID122, Heap-based Buffer Overflow, where the commercial tool scores a few percents higher, and CWE-ID590, Free of Memory not on the Heap.

These results suggest that our representation has good prospects to be used in vulnerability discovery tools. For almost every vulnerability type our prediction accuracy performance is better than 80\% and for many it is higher than 90\%.

\section{Discussion and future work}

Software in production is usually complex and large, capable of performing many different functions in different use cases.
On the contrary, programs in our evaluation datasets are single-purpose, solving a single task with a relatively small number of steps.
Additionally, the entirety of each program in Juliet test suite is relevant to vulnerability discovery tasks, unlike real software where most of the code is not vulnerable and only a small part of it may have an issue. 
This can potentially be solved by introducing representations that can be computed on different levels of coarseness. 
This is a non-trivial task, but our findings hint that once completed we may be able to achieve far better results for different problems on production software than is currently possible. 
Additionally, we need to get a better understanding of what properties are captured with such a representation and how is best to use those or how to add other desirable properties. 
Another challenge left for future work is extending this approach to cross-architecture and cross-compiler binaries. 

There are several avenues for extending our work. First, it will be interesting to see whether using recent extensions of GCNs, such as the MixHop model~\cite{sami} that propagates information through higher-order node neighbourhoods, will result in better performance. Additionally, to test the utility of Bin2Vec in real-world problems, we would like to apply it  to analyze more complex and larger-scale vulnerability datasets.

\section{Conclusion}

We introduced Bin2Vec, a new model for learning distributed representations of binary executable programs. Our learned representation has strong potential to be used in the context of a wide variety of binary analysis tasks. We demonstrate this by putting our learned representations to use for classification in two semantically different tasks - algorithm classification and vulnerability discovery. We show that for both tasks our proposed representation achieves better qualitative and quantitative performance in comparison to state-of-the-art approaches, including inst2vec and common machine learning baselines. 

\section*{Availability of data and materials}
All of the data used in this study is publicly available.

\section*{Funding}
The authors are grateful for the funding to Information Sciences Institute of University of Southern California.

\bibliographystyle{spbasic}
\bibliography{bibliography} 

\begin{thebibliography}{55}
\providecommand{\natexlab}[1]{#1}
\providecommand{\url}[1]{{#1}}
\providecommand{\urlprefix}{URL }
\expandafter\ifx\csname urlstyle\endcsname\relax
  \providecommand{\doi}[1]{DOI~\discretionary{}{}{}#1}\else
  \providecommand{\doi}{DOI~\discretionary{}{}{}\begingroup
  \urlstyle{rm}\Url}\fi
\providecommand{\eprint}[2][]{\url{#2}}

\bibitem[{Aafer et~al.(2013)Aafer, Du, and
  Yin}]{DBLP:conf/securecomm/AaferDY13}
Aafer Y, Du W, Yin H (2013) Droidapiminer: Mining api-level features for robust
  malware detection in android. In: Zia TA, Zomaya AY, Varadharajan V, Mao ZM
  (eds) Security and Privacy in Communication Networks - 9th International
  {ICST} Conference, SecureComm 2013, Sydney, NSW, Australia, September 25-28,
  2013, Revised Selected Papers, Springer, vol 127, pp 86--103,
  \doi{10.1007/978-3-319-04283-1\_6},
  \urlprefix\url{https://doi.org/10.1007/978-3-319-04283-1\_6}

\bibitem[{Abu-El-Haija et~al.(2019)Abu-El-Haija, Perozzi, Kapoor, Alipourfard,
  Lerman, Harutyunyan, Steeg, and Galstyan}]{sami}
Abu-El-Haija S, Perozzi B, Kapoor A, Alipourfard N, Lerman K, Harutyunyan H,
  Steeg GV, Galstyan A (2019) {M}ix{H}op: Higher-order graph convolutional
  architectures via sparsified neighborhood mixing. PMLR, Long Beach,
  California, USA, Proceedings of Machine Learning Research, vol~97, pp 21--29

\bibitem[{Allamanis et~al.(2018)Allamanis, Barr, Devanbu, and
  Sutton}]{DBLP:journals/csur/AllamanisBDS18}
Allamanis M, Barr ET, Devanbu PT, Sutton CA (2018) A survey of machine learning
  for big code and naturalness. {ACM} Comput Surv 51(4):81:1--81:37,
  \doi{10.1145/3212695}, \urlprefix\url{https://doi.org/10.1145/3212695}

\bibitem[{Andriesse et~al.(2016)Andriesse, Chen, van~der Veen, Slowinska, and
  Bos}]{victor}
Andriesse D, Chen X, van~der Veen V, Slowinska A, Bos H (2016) An in-depth
  analysis of disassembly on full-scale x86/x64 binaries. In: {USENIX},
  {USENIX} Association, Austin, TX,
  \urlprefix\url{https://www.usenix.org/conference/usenixsecurity16/technical-sessions/presentation/andriesse}

\bibitem[{Ben{-}Nun et~al.(2018)Ben{-}Nun, Jakobovits, and
  Hoefler}]{DBLP:conf/nips/Ben-NunJH18}
Ben{-}Nun T, Jakobovits AS, Hoefler T (2018) Neural code comprehension: {A}
  learnable representation of code semantics. In: Bengio S, Wallach HM,
  Larochelle H, Grauman K, Cesa{-}Bianchi N, Garnett R (eds) Advances in Neural
  Information Processing Systems 31: Annual Conference on Neural Information
  Processing Systems 2018, NeurIPS 2018, 3-8 December 2018, Montr{\'{e}}al,
  Canada., pp 3589--3601,
  \urlprefix\url{http://papers.nips.cc/paper/7617-neural-code-comprehension-a-learnable-representation-of-code-semantics}

\bibitem[{Bengio and LeCun(2017)}]{DBLP:conf/iclr/2017}
Bengio Y, LeCun Y (eds) (2017) 5th International Conference on Learning
  Representations, {ICLR} 2017, Toulon, France, April 24-26, 2017, Conference
  Track Proceedings, OpenReview.net,
  \urlprefix\url{https://openreview.net/group?id=ICLR.cc/2017/conference}

\bibitem[{Boland and Black(2012)}]{DBLP:journals/computer/BolandB12}
Boland T, Black PE (2012) Juliet 1.1 {C/C++} and java test suite. {IEEE}
  Computer 45(10):88--90, \doi{10.1109/MC.2012.345},
  \urlprefix\url{https://doi.org/10.1109/MC.2012.345}

\bibitem[{Cha et~al.(2012)Cha, Avgerinos, Rebert, and
  Brumley}]{DBLP:conf/sp/ChaARB12}
Cha SK, Avgerinos T, Rebert A, Brumley D (2012) Unleashing mayhem on binary
  code. In: {IEEE} Symposium on Security and Privacy, {SP} 2012, 21-23 May
  2012, San Francisco, California, {USA}, {IEEE} Computer Society, pp 380--394,
  \doi{10.1109/SP.2012.31}, \urlprefix\url{https://doi.org/10.1109/SP.2012.31}

\bibitem[{Chen et~al.(2018)Chen, Ma, and Xiao}]{DBLP:conf/iclr/ChenMX18}
Chen J, Ma T, Xiao C (2018) Fastgcn: Fast learning with graph convolutional
  networks via importance sampling. In: 6th International Conference on
  Learning Representations, {ICLR} 2018, Vancouver, BC, Canada, April 30 - May
  3, 2018, Conference Track Proceedings, OpenReview.net,
  \urlprefix\url{https://openreview.net/forum?id=rytstxWAW}

\bibitem[{Ding et~al.(2019)Ding, Fung, and Charland}]{DBLP:conf/sp/DingFC19}
Ding SHH, Fung BCM, Charland P (2019) Asm2vec: Boosting static representation
  robustness for binary clone search against code obfuscation and compiler
  optimization. In: 2019 {IEEE} Symposium on Security and Privacy, {SP} 2019,
  San Francisco, CA, USA, May 19-23, 2019, {IEEE}, pp 472--489,
  \doi{10.1109/SP.2019.00003},
  \urlprefix\url{https://doi.org/10.1109/SP.2019.00003}

\bibitem[{Eschweiler et~al.(2016)Eschweiler, Yakdan, and
  Gerhards{-}Padilla}]{DBLP:conf/ndss/EschweilerYG16}
Eschweiler S, Yakdan K, Gerhards{-}Padilla E (2016) discovre: Efficient
  cross-architecture identification of bugs in binary code. In: 23rd Annual
  Network and Distributed System Security Symposium, {NDSS} 2016, San Diego,
  California, USA, February 21-24, 2016, The Internet Society,
  \urlprefix\url{http://wp.internetsociety.org/ndss/wp-content/uploads/sites/25/2017/09/discovre-efficient-cross-architecture-identification-bugs-binary-code.pdf}

\bibitem[{Feng et~al.(2016)Feng, Zhou, Xu, Cheng, Testa, and
  Yin}]{DBLP:conf/ccs/FengZXCTY16}
Feng Q, Zhou R, Xu C, Cheng Y, Testa B, Yin H (2016) Scalable graph-based bug
  search for firmware images. In: Weippl ER, Katzenbeisser S, Kruegel C, Myers
  AC, Halevi S (eds) Proceedings of the 2016 {ACM} {SIGSAC} Conference on
  Computer and Communications Security, Vienna, Austria, October 24-28, 2016,
  {ACM}, pp 480--491, \doi{10.1145/2976749.2978370},
  \urlprefix\url{https://doi.org/10.1145/2976749.2978370}

\bibitem[{Goseva{-}Popstojanova and
  Perhinschi(2015)}]{DBLP:journals/infsof/Goseva-Popstojanova15}
Goseva{-}Popstojanova K, Perhinschi A (2015) On the capability of static code
  analysis to detect security vulnerabilities. Information {\&} Software
  Technology 68:18--33, \doi{10.1016/j.infsof.2015.08.002},
  \urlprefix\url{https://doi.org/10.1016/j.infsof.2015.08.002}

\bibitem[{Grieco et~al.(2016)Grieco, Grinblat, Uzal, Rawat, Feist, and
  Mounier}]{DBLP:conf/codaspy/GriecoGURFM16}
Grieco G, Grinblat GL, Uzal LC, Rawat S, Feist J, Mounier L (2016) Toward
  large-scale vulnerability discovery using machine learning. In: Bertino E,
  Sandhu R, Pretschner A (eds) Proceedings of the Sixth {ACM} on Conference on
  Data and Application Security and Privacy, {CODASPY} 2016, New Orleans, LA,
  USA, March 9-11, 2016, {ACM}, pp 85--96, \doi{10.1145/2857705.2857720},
  \urlprefix\url{https://doi.org/10.1145/2857705.2857720}

\bibitem[{Guyon et~al.(2017)Guyon, von Luxburg, Bengio, Wallach, Fergus,
  Vishwanathan, and Garnett}]{DBLP:conf/nips/2017}
Guyon I, von Luxburg U, Bengio S, Wallach HM, Fergus R, Vishwanathan SVN,
  Garnett R (eds) (2017) Advances in Neural Information Processing Systems 30:
  Annual Conference on Neural Information Processing Systems 2017, 4-9 December
  2017, Long Beach, CA, {USA}

\bibitem[{Hamilton et~al.(2017)Hamilton, Ying, and
  Leskovec}]{DBLP:conf/nips/HamiltonYL17}
Hamilton WL, Ying Z, Leskovec J (2017) Inductive representation learning on
  large graphs. In:  \cite{DBLP:conf/nips/2017}, pp 1024--1034,
  \urlprefix\url{http://papers.nips.cc/paper/6703-inductive-representation-learning-on-large-graphs}

\bibitem[{He et~al.(2018)He, Ivanov, Tsankov, Raychev, and
  Vechev}]{DBLP:conf/ccs/HeITRV18}
He J, Ivanov P, Tsankov P, Raychev V, Vechev MT (2018) Debin: Predicting debug
  information in stripped binaries. In: Lie D, Mannan M, Backes M, Wang X (eds)
  Proceedings of the 2018 {ACM} {SIGSAC} Conference on Computer and
  Communications Security, {CCS} 2018, Toronto, ON, Canada, October 15-19,
  2018, {ACM}, pp 1667--1680, \doi{10.1145/3243734.3243866},
  \urlprefix\url{https://doi.org/10.1145/3243734.3243866}

\bibitem[{Hindle et~al.(2016)Hindle, Barr, Gabel, Su, and
  Devanbu}]{DBLP:journals/cacm/HindleBGS16}
Hindle A, Barr ET, Gabel M, Su Z, Devanbu PT (2016) On the naturalness of
  software. Commun {ACM} 59(5):122--131, \doi{10.1145/2902362},
  \urlprefix\url{https://doi.org/10.1145/2902362}

\bibitem[{Kang et~al.(2016)Kang, Yerima, Sezer, and
  McLaughlin}]{DBLP:journals/ijcysa/KangYSM16}
Kang B, Yerima SY, Sezer S, McLaughlin K (2016) N-gram opcode analysis for
  android malware detection. {IJCSA} 1(1):231--255,
  \doi{10.22619/ijcsa.2016.1001011},
  \urlprefix\url{https://doi.org/10.22619/ijcsa.2016.1001011}

\bibitem[{Karbab et~al.(2018)Karbab, Debbabi, Derhab, and
  Mouheb}]{DBLP:journals/di/KarbabDDM18}
Karbab EB, Debbabi M, Derhab A, Mouheb D (2018) Maldozer: Automatic framework
  for android malware detection using deep learning. Digital Investigation
  24:S48--S59, \doi{10.1016/j.diin.2018.01.007},
  \urlprefix\url{https://doi.org/10.1016/j.diin.2018.01.007}

\bibitem[{Kipf and Welling(2017)}]{DBLP:conf/iclr/KipfW17}
Kipf TN, Welling M (2017) Semi-supervised classification with graph
  convolutional networks. In:  \cite{DBLP:conf/iclr/2017},
  \urlprefix\url{https://openreview.net/forum?id=SJU4ayYgl}

\bibitem[{Kolosnjaji et~al.(2016)Kolosnjaji, Zarras, Webster, and
  Eckert}]{DBLP:conf/ausai/KolosnjajiZWE16}
Kolosnjaji B, Zarras A, Webster GD, Eckert C (2016) Deep learning for
  classification of malware system call sequences. In: Kang BH, Bai Q (eds)
  {AI} 2016: Advances in Artificial Intelligence - 29th Australasian Joint
  Conference, Hobart, TAS, Australia, December 5-8, 2016, Proceedings,
  Springer, Lecture Notes in Computer Science, vol 9992, pp 137--149,
  \doi{10.1007/978-3-319-50127-7\_11},
  \urlprefix\url{https://doi.org/10.1007/978-3-319-50127-7\_11}

\bibitem[{Lee et~al.(2018)Lee, Choi, Shin, and
  Kwak}]{DBLP:journals/tjs/LeeCSK18}
Lee T, Choi B, Shin Y, Kwak J (2018) Automatic malware mutant detection and
  group classification based on the n-gram and clustering coefficient. The
  Journal of Supercomputing 74(8):3489--3503, \doi{10.1007/s11227-015-1594-6},
  \urlprefix\url{https://doi.org/10.1007/s11227-015-1594-6}

\bibitem[{Li et~al.(2019{\natexlab{a}})Li, Zhang, Yao, and
  Yin}]{DBLP:conf/ict/LiZYY19}
Li B, Zhang Y, Yao J, Yin T (2019{\natexlab{a}}) {MDBA:} detecting malware
  based on bytes n-gram with association mining. In: 26th International
  Conference on Telecommunications, {ICT} 2019, Hanoi, Vietnam, April 8-10,
  2019, {IEEE}, pp 227--232, \doi{10.1109/ICT.2019.8798828},
  \urlprefix\url{https://doi.org/10.1109/ICT.2019.8798828}

\bibitem[{Li et~al.(2016)Li, Tarlow, Brockschmidt, and
  Zemel}]{DBLP:journals/corr/LiTBZ15}
Li Y, Tarlow D, Brockschmidt M, Zemel RS (2016) Gated graph sequence neural
  networks. In: Bengio Y, LeCun Y (eds) 4th International Conference on
  Learning Representations, {ICLR} 2016, San Juan, Puerto Rico, May 2-4, 2016,
  Conference Track Proceedings, \urlprefix\url{http://arxiv.org/abs/1511.05493}

\bibitem[{Li et~al.(2019{\natexlab{b}})Li, Gu, Dullien, Vinyals, and
  Kohli}]{DBLP:conf/icml/LiGDVK19}
Li Y, Gu C, Dullien T, Vinyals O, Kohli P (2019{\natexlab{b}}) Graph matching
  networks for learning the similarity of graph structured objects. In:
  Chaudhuri K, Salakhutdinov R (eds) Proceedings of the 36th International
  Conference on Machine Learning, {ICML} 2019, 9-15 June 2019, Long Beach,
  California, {USA}, {PMLR}, Proceedings of Machine Learning Research, vol~97,
  pp 3835--3845, \urlprefix\url{http://proceedings.mlr.press/v97/li19d.html}

\bibitem[{Li et~al.(2018)Li, Zou, Xu, Ou, Jin, Wang, Deng, and
  Zhong}]{DBLP:conf/ndss/LiZXO0WDZ18}
Li Z, Zou D, Xu S, Ou X, Jin H, Wang S, Deng Z, Zhong Y (2018) Vuldeepecker:
  {A} deep learning-based system for vulnerability detection. In: 25th Annual
  Network and Distributed System Security Symposium, {NDSS} 2018, San Diego,
  California, USA, February 18-21, 2018, The Internet Society,
  \urlprefix\url{http://wp.internetsociety.org/ndss/wp-content/uploads/sites/25/2018/02/ndss2018\_03A-2\_Li\_paper.pdf}

\bibitem[{Lin et~al.(2017{\natexlab{a}})Lin, Zhang, Luo, Pan, and
  Xiang}]{DBLP:conf/ccs/LinZLPX17}
Lin G, Zhang J, Luo W, Pan L, Xiang Y (2017{\natexlab{a}}) {POSTER:}
  vulnerability discovery with function representation learning from unlabeled
  projects. In:  \cite{DBLP:conf/ccs/2017}, pp 2539--2541,
  \doi{10.1145/3133956.3138840},
  \urlprefix\url{https://doi.org/10.1145/3133956.3138840}

\bibitem[{Lin et~al.(2017{\natexlab{b}})Lin, Feng, dos Santos, Yu, Xiang, Zhou,
  and Bengio}]{DBLP:conf/iclr/LinFSYXZB17}
Lin Z, Feng M, dos Santos CN, Yu M, Xiang B, Zhou B, Bengio Y
  (2017{\natexlab{b}}) A structured self-attentive sentence embedding. In:
  \cite{DBLP:conf/iclr/2017},
  \urlprefix\url{https://openreview.net/forum?id=BJC\_jUqxe}

\bibitem[{Liu et~al.(2019)Liu, Chen, Li, Zhou, Li, Song, and
  Qi}]{DBLP:conf/aaai/LiuCLZLSQ19}
Liu Z, Chen C, Li L, Zhou J, Li X, Song L, Qi Y (2019) Geniepath: Graph neural
  networks with adaptive receptive paths. In: The Thirty-Third {AAAI}
  Conference on Artificial Intelligence, {AAAI} 2019, The Thirty-First
  Innovative Applications of Artificial Intelligence Conference, {IAAI} 2019,
  The Ninth {AAAI} Symposium on Educational Advances in Artificial
  Intelligence, {EAAI} 2019, Honolulu, Hawaii, USA, January 27 - February 1,
  2019, {AAAI} Press, pp 4424--4431, \doi{10.1609/aaai.v33i01.33014424},
  \urlprefix\url{https://doi.org/10.1609/aaai.v33i01.33014424}

\bibitem[{Mikolov et~al.(2013{\natexlab{a}})Mikolov, Chen, Corrado, and
  Dean}]{DBLP:journals/corr/abs-1301-3781}
Mikolov T, Chen K, Corrado G, Dean J (2013{\natexlab{a}}) Efficient estimation
  of word representations in vector space. In: Bengio Y, LeCun Y (eds) 1st
  International Conference on Learning Representations, {ICLR} 2013,
  Scottsdale, Arizona, USA, May 2-4, 2013, Workshop Track Proceedings,
  \urlprefix\url{http://arxiv.org/abs/1301.3781}

\bibitem[{Mikolov et~al.(2013{\natexlab{b}})Mikolov, Sutskever, Chen, Corrado,
  and Dean}]{DBLP:conf/nips/MikolovSCCD13}
Mikolov T, Sutskever I, Chen K, Corrado GS, Dean J (2013{\natexlab{b}})
  Distributed representations of words and phrases and their compositionality.
  In: Burges CJC, Bottou L, Ghahramani Z, Weinberger KQ (eds) Advances in
  Neural Information Processing Systems 26: 27th Annual Conference on Neural
  Information Processing Systems 2013. Proceedings of a meeting held December
  5-8, 2013, Lake Tahoe, Nevada, United States., pp 3111--3119,
  \urlprefix\url{http://papers.nips.cc/paper/5021-distributed-representations-of-words-and-phrases-and-their-compositionality}

\bibitem[{Monti et~al.(2017)Monti, Boscaini, Masci, Rodol{\`{a}}, Svoboda, and
  Bronstein}]{DBLP:conf/cvpr/MontiBMRSB17}
Monti F, Boscaini D, Masci J, Rodol{\`{a}} E, Svoboda J, Bronstein MM (2017)
  Geometric deep learning on graphs and manifolds using mixture model cnns. In:
  2017 {IEEE} Conference on Computer Vision and Pattern Recognition, {CVPR}
  2017, Honolulu, HI, USA, July 21-26, 2017, {IEEE} Computer Society, pp
  5425--5434, \doi{10.1109/CVPR.2017.576},
  \urlprefix\url{https://doi.org/10.1109/CVPR.2017.576}

\bibitem[{Mou et~al.(2016)Mou, Li, Zhang, Wang, and
  Jin}]{DBLP:conf/aaai/MouLZWJ16}
Mou L, Li G, Zhang L, Wang T, Jin Z (2016) Convolutional neural networks over
  tree structures for programming language processing. In: Schuurmans D,
  Wellman MP (eds) Proceedings of the Thirtieth {AAAI} Conference on Artificial
  Intelligence, February 12-17, {AAAI} Press, 2016, Phoenix, Arizona, {USA.},
  pp 1287--1293,
  \urlprefix\url{http://www.aaai.org/ocs/index.php/AAAI/AAAI16/paper/view/11775}

\bibitem[{Murtaza et~al.(2016)Murtaza, Khreich, Hamou{-}Lhadj, and
  Bener}]{DBLP:journals/jss/MurtazaKHB16}
Murtaza SS, Khreich W, Hamou{-}Lhadj A, Bener AB (2016) Mining trends and
  patterns of software vulnerabilities. Journal of Systems and Software
  117:218--228, \doi{10.1016/j.jss.2016.02.048},
  \urlprefix\url{https://doi.org/10.1016/j.jss.2016.02.048}

\bibitem[{Pang et~al.(2015)Pang, Xue, and Namin}]{DBLP:conf/icmla/PangXN15}
Pang Y, Xue X, Namin AS (2015) Predicting vulnerable software components
  through n-gram analysis and statistical feature selection. In: Li T, Kurgan
  LA, Palade V, Goebel R, Holzinger A, Verspoor K, Wani MA (eds) 14th {IEEE}
  International Conference on Machine Learning and Applications, {ICMLA} 2015,
  Miami, FL, USA, December 9-11, 2015, {IEEE}, pp 543--548,
  \doi{10.1109/ICMLA.2015.99},
  \urlprefix\url{https://doi.org/10.1109/ICMLA.2015.99}

\bibitem[{Perozzi et~al.(2014)Perozzi, Al{-}Rfou, and
  Skiena}]{DBLP:conf/kdd/PerozziAS14}
Perozzi B, Al{-}Rfou R, Skiena S (2014) Deepwalk: online learning of social
  representations. In: Macskassy SA, Perlich C, Leskovec J, Wang W, Ghani R
  (eds) The 20th {ACM} {SIGKDD} International Conference on Knowledge Discovery
  and Data Mining, {KDD} '14, New York, NY, {USA} - August 24 - 27, 2014,
  {ACM}, pp 701--710, \doi{10.1145/2623330.2623732},
  \urlprefix\url{https://doi.org/10.1145/2623330.2623732}

\bibitem[{Rawat and Mounier(2012)}]{DBLP:conf/ssiri/RawatM12}
Rawat S, Mounier L (2012) Finding buffer overflow inducing loops in binary
  executables. In: Sixth International Conference on Software Security and
  Reliability, {SERE} 2012, Gaithersburg, Maryland, USA, 20-22 June 2012,
  {IEEE}, pp 177--186, \doi{10.1109/SERE.2012.30},
  \urlprefix\url{https://doi.org/10.1109/SERE.2012.30}

\bibitem[{Ray et~al.(2016)Ray, Hellendoorn, Godhane, Tu, Bacchelli, and
  Devanbu}]{DBLP:conf/icse/RayHGTBD16}
Ray B, Hellendoorn V, Godhane S, Tu Z, Bacchelli A, Devanbu PT (2016) On the
  "naturalness" of buggy code. In: Dillon LK, Visser W, Williams L (eds)
  Proceedings of the 38th International Conference on Software Engineering,
  {ICSE} 2016, Austin, TX, USA, May 14-22, 2016, {ACM}, pp 428--439,
  \doi{10.1145/2884781.2884848},
  \urlprefix\url{https://doi.org/10.1145/2884781.2884848}

\bibitem[{Santos et~al.(2009)Santos, Penya, Devesa, and
  Bringas}]{DBLP:conf/iceis/SantosPDB09}
Santos I, Penya YK, Devesa J, Bringas PG (2009) N-grams-based file signatures
  for malware detection. In: Cordeiro J, Filipe J (eds) {ICEIS} 2009 -
  Proceedings of the 11th International Conference on Enterprise Information
  Systems, Volume AIDSS, Milan, Italy, May 6-10, 2009, pp 317--320

\bibitem[{Shoshitaishvili et~al.(2016)Shoshitaishvili, Wang, Salls, Stephens,
  Polino, Dutcher, Grosen, Feng, Hauser, Kr{\"{u}}gel, and
  Vigna}]{DBLP:conf/sp/Shoshitaishvili16}
Shoshitaishvili Y, Wang R, Salls C, Stephens N, Polino M, Dutcher A, Grosen J,
  Feng S, Hauser C, Kr{\"{u}}gel C, Vigna G (2016) {SOK:} (state of) the art of
  war: Offensive techniques in binary analysis. In: {IEEE} Symposium on
  Security and Privacy, {SP} 2016, San Jose, CA, USA, May 22-26, 2016, {IEEE}
  Computer Society, pp 138--157, \doi{10.1109/SP.2016.17},
  \urlprefix\url{https://doi.org/10.1109/SP.2016.17}

\bibitem[{Simonyan and Zisserman(2015)}]{DBLP:journals/corr/SimonyanZ14a}
Simonyan K, Zisserman A (2015) Very deep convolutional networks for large-scale
  image recognition. In: Bengio Y, LeCun Y (eds) 3rd International Conference
  on Learning Representations, {ICLR} 2015, San Diego, CA, USA, May 7-9, 2015,
  Conference Track Proceedings, \urlprefix\url{http://arxiv.org/abs/1409.1556}

\bibitem[{Theisen et~al.(2015)Theisen, Herzig, Morrison, Murphy, and
  Williams}]{DBLP:conf/icse/TheisenHMMW15}
Theisen C, Herzig K, Morrison P, Murphy B, Williams LA (2015) Approximating
  attack surfaces with stack traces. In: Bertolino A, Canfora G, Elbaum SG
  (eds) 37th {IEEE/ACM} International Conference on Software Engineering,
  {ICSE} 2015, Florence, Italy, May 16-24, 2015, Volume 2, {IEEE} Computer
  Society, pp 199--208, \doi{10.1109/ICSE.2015.148},
  \urlprefix\url{https://doi.org/10.1109/ICSE.2015.148}

\bibitem[{Thuraisingham et~al.(2017)Thuraisingham, Evans, Malkin, and
  Xu}]{DBLP:conf/ccs/2017}
Thuraisingham BM, Evans D, Malkin T, Xu D (eds) (2017) Proceedings of the 2017
  {ACM} {SIGSAC} Conference on Computer and Communications Security, {CCS}
  2017, Dallas, TX, USA, October 30 - November 03, 2017, {ACM},
  \doi{10.1145/3133956}, \urlprefix\url{https://doi.org/10.1145/3133956}

\bibitem[{Velicheti et~al.(2014)Velicheti, Feiock, Peiris, Raje, and
  Hill}]{DBLP:conf/csiirw/VelichetiFPRH14}
Velicheti LMR, Feiock DC, Peiris M, Raje RR, Hill JH (2014) Towards modeling
  the behavior of static code analysis tools. In: Abercrombie RK, McDonald JT
  (eds) Cyber and Information Security Research Conference, {CISR} '14, Oak
  Ridge, TN, USA, April 8-10, 2014, {ACM}, pp 17--20,
  \doi{10.1145/2602087.2602101},
  \urlprefix\url{https://doi.org/10.1145/2602087.2602101}

\bibitem[{Velickovic et~al.(2017)Velickovic, Cucurull, Casanova, Romero,
  Li{\`{o}}, and Bengio}]{DBLP:journals/corr/abs-1710-10903}
Velickovic P, Cucurull G, Casanova A, Romero A, Li{\`{o}} P, Bengio Y (2017)
  Graph attention networks. CoRR abs/1710.10903,
  \urlprefix\url{http://arxiv.org/abs/1710.10903}, \eprint{1710.10903}

\bibitem[{White et~al.(2016)White, Tufano, Vendome, and
  Poshyvanyk}]{DBLP:conf/kbse/WhiteTVP16}
White M, Tufano M, Vendome C, Poshyvanyk D (2016) Deep learning code fragments
  for code clone detection. In: Lo D, Apel S, Khurshid S (eds) Proceedings of
  the 31st {IEEE/ACM} International Conference on Automated Software
  Engineering, {ASE} 2016, Singapore, September 3-7, 2016, {ACM}, pp 87--98,
  \doi{10.1145/2970276.2970326},
  \urlprefix\url{https://doi.org/10.1145/2970276.2970326}

\bibitem[{Wu et~al.(2016)Wu, Wang, Li, and
  Zhang}]{DBLP:journals/infsof/WuWLZ16}
Wu S, Wang P, Li X, Zhang Y (2016) Effective detection of android malware based
  on the usage of data flow apis and machine learning. Information {\&}
  Software Technology 75:17--25, \doi{10.1016/j.infsof.2016.03.004},
  \urlprefix\url{https://doi.org/10.1016/j.infsof.2016.03.004}

\bibitem[{W{\"{u}}chner et~al.(2015)W{\"{u}}chner, Ochoa, and
  Pretschner}]{DBLP:conf/dimva/WuchnerOP15}
W{\"{u}}chner T, Ochoa M, Pretschner A (2015) Robust and effective malware
  detection through quantitative data flow graph metrics. In: Almgren M,
  Gulisano V, Maggi F (eds) Detection of Intrusions and Malware, and
  Vulnerability Assessment - 12th International Conference, {DIMVA} 2015,
  Milan, Italy, July 9-10, 2015, Proceedings, Springer, Lecture Notes in
  Computer Science, vol 9148, pp 98--118, \doi{10.1007/978-3-319-20550-2\_6},
  \urlprefix\url{https://doi.org/10.1007/978-3-319-20550-2\_6}

\bibitem[{Xu et~al.(2017)Xu, Liu, Feng, Yin, Song, and
  Song}]{DBLP:conf/ccs/XuLFYSS17}
Xu X, Liu C, Feng Q, Yin H, Song L, Song D (2017) Neural network-based graph
  embedding for cross-platform binary code similarity detection. In:
  \cite{DBLP:conf/ccs/2017}, pp 363--376, \doi{10.1145/3133956.3134018},
  \urlprefix\url{https://doi.org/10.1145/3133956.3134018}

\bibitem[{Yamaguchi et~al.(2014)Yamaguchi, Golde, Arp, and
  Rieck}]{DBLP:conf/sp/YamaguchiGAR14}
Yamaguchi F, Golde N, Arp D, Rieck K (2014) Modeling and discovering
  vulnerabilities with code property graphs. In: 2014 {IEEE} Symposium on
  Security and Privacy, {SP} 2014, Berkeley, CA, USA, May 18-21, 2014, {IEEE}
  Computer Society, pp 590--604, \doi{10.1109/SP.2014.44},
  \urlprefix\url{https://doi.org/10.1109/SP.2014.44}

\bibitem[{Zalewski(2017)}]{afl}
Zalewski M (2017) {American Fuzzy Lop}. \url{http://lcamtuf.coredump.cx/afl/}

\bibitem[{Zhang et~al.(2017)Zhang, Shen, Wang, Gan, Henao, and
  Carin}]{DBLP:conf/nips/ZhangSWGHC17}
Zhang Y, Shen D, Wang G, Gan Z, Henao R, Carin L (2017) Deconvolutional
  paragraph representation learning. In:  \cite{DBLP:conf/nips/2017}, pp
  4169--4179,
  \urlprefix\url{http://papers.nips.cc/paper/7005-deconvolutional-paragraph-representation-learning}

\bibitem[{Zhou et~al.(2019)Zhou, Liu, Siow, Du, and
  Liu}]{DBLP:conf/nips/ZhouLSD019}
Zhou Y, Liu S, Siow JK, Du X, Liu Y (2019) Devign: Effective vulnerability
  identification by learning comprehensive program semantics via graph neural
  networks. In: Wallach HM, Larochelle H, Beygelzimer A, d'Alch{\'{e}}{-}Buc F,
  Fox EB, Garnett R (eds) Advances in Neural Information Processing Systems 32:
  Annual Conference on Neural Information Processing Systems 2019, NeurIPS
  2019, 8-14 December 2019, Vancouver, BC, Canada, pp 10197--10207,
  \urlprefix\url{http://papers.nips.cc/paper/9209-devign-effective-vulnerability-identification-by-learning-comprehensive-program-semantics-via-graph-neural-networks}

\bibitem[{Zuo et~al.(2019)Zuo, Li, Young, Luo, Zeng, and
  Zhang}]{DBLP:conf/ndss/ZuoLYL0Z19}
Zuo F, Li X, Young P, Luo L, Zeng Q, Zhang Z (2019) Neural machine translation
  inspired binary code similarity comparison beyond function pairs. In: 26th
  Annual Network and Distributed System Security Symposium, {NDSS} 2019, San
  Diego, California, USA, February 24-27, 2019, The Internet Society,
  \urlprefix\url{https://www.ndss-symposium.org/ndss-paper/neural-machine-translation-inspired-binary-code-similarity-comparison-beyond-function-pairs/}

\end{thebibliography}
\end{document}